\documentclass[amsmath,amssymb,amsfonts,reprint,floatfix,notitlepage,tightenlines]{revtex4-1}
\pdfoutput=1
\usepackage{graphicx}
\usepackage{dcolumn}
\usepackage{bm}

\usepackage[english]{babel}
\makeatletter
\def\bbl@set@language#1{%
  \edef\languagename{%
    \ifnum\escapechar=\expandafter`\string#1\@empty
    \else\string#1\@empty\fi}%
  \@ifundefined{babel@language@alias@\languagename}{}{%
    \edef\languagename{\@nameuse{babel@language@alias@\languagename}}%
  }
  \select@language{\languagename}%
  \expandafter\ifx\csname date\languagename\endcsname\relax\else
    \if@filesw
      \protected@write\@auxout{}{\string\select@language{\languagename}}%
      \bbl@for\bbl@tempa\BabelContentsFiles{%
        \addtocontents{\bbl@tempa}{\xstring\select@language{\languagename}}}%
      \bbl@usehooks{write}{}%
    \fi
  \fi}
\newcommand{\DeclareLanguageAlias}[2]{\global\@namedef{babel@language@alias@#1}{#2}}
\makeatother
\DeclareLanguageAlias{en}{english}
\begin{document}
\title{Fundamental limits to signal integrity in nonlinear parametric optical circulators}
\author{Ian A. D. Williamson}
\email{ian.williamson@utexas.edu}
\author{Zheng Wang}
\email{zheng.wang@austin.utexas.edu}
\affiliation{Microelectronics Research Center, The University of Texas at Austin, Austin, TX 78758 USA}
%\keywords{(060.5625) Radio frequency photonics; (070.4340) Nonlinear optical signal processing; (190.4390) Nonlinear optics, integrated optics; (190.4410) Nonlinear optics, parametric processes; (230.3240) Isolators.}
\begin{abstract}
We characterize the response of a parametric nonlinear optical circulator to realistic signals that have finite bandwidths. Our results show that intermodulation distortion (IMD), rather than pump depletion or compression, limits the maximal operating signal power and the dynamic range of nonlinear parametric circulators. This limitation holds even in the undepleted pump regime where nonlinear circulators are not constrained by dynamic reciprocity. With a realistic pump power, noise floor, and nonlinear waveguide, our numerical modeling demonstrates a maximally achievable spur-free dynamic range (SFDR) of 81 dB.
\end{abstract}

\maketitle

Circulators play the key role of separating high-power outgoing signals from low-power received signals in optical transceivers and interferometers for communication, signal-processing, ranging, and imaging applications in optics and microwave photonics. Because these applications involve broadband signals, the linearity of the circulator transfer function is crucial for maintaining a large signal-to-noise ratio (SNR), especially for systems requiring high spectral efficiency or digital modulation schemes with high peak-to-average power ratios. High linearity of the circulator transfer function in the forward direction minimizes signal distortion and the associated spurious signals that cause \textit{inter-channel} and \textit{intra-channel} interference \cite{marpaung_integrated_2013}.

Conventionally, in order to break reciprocity, optical circulators rely on the gyromagnetic effect, a linear magneto optical (MO) effect \cite{bi_-chip_2011, shoji_optical_2016}. However MO materials are challenging for on-chip integration due to material incompatibility as well as the requirement of large interaction lengths that overcome weak magnetic effects at optical frequencies. More recently, time-reversal symmetry breaking has been achieved via second-order \cite{rangelov_nonlinear_2017, wang_non-reciprocal_2017} and third-order optical nonlinearities \cite{fan_all-silicon_2012, peng_paritytime-symmetric_2014, hua_demonstration_2016} as well as stimulated Brillouin scattering (SBS) \cite{huang_complete_2011}. These approaches show great promise for realizing CMOS compatible circulators with smaller device sizes and improved integration. However, dynamic reciprocity limits a nonlinear circulator's ability to isolate signals in neighboring frequency bands \cite{shi_limitations_2015}. This makes it challenging to use such circulator designs in broadband systems where signals at multiple ports must be simultaneously routed. Moreover, breaking time-reversal symmetry with nonlinearity imposes a fundamental lower limit on the operating signal power level, below which the system scatters light reciprocally.

A recently proposed class of nonlinear circulator and isolator designs have operated in the \textit{undepleted pump} regime of parametric three-wave mixing. In these systems the interaction of signal and idler waves is effectively linear, overcoming the constraint of dynamic reciprocity while, in principal, eliminating the lower limit on operational signal power. However, the overall design space for parametric nonlinear circulators has not been fully explored, and the range of allowable signal powers has not been quantified. Moreover, the response these devices to finite-bandwidth signals has not been considered.  It is therefore critical to evaluate the suitability of parametric nonlinear processes in realistic circulator designs, which could be constrained by practical limits on pump power and noise performance.

\begin{figure}
  \centering
  \includegraphics{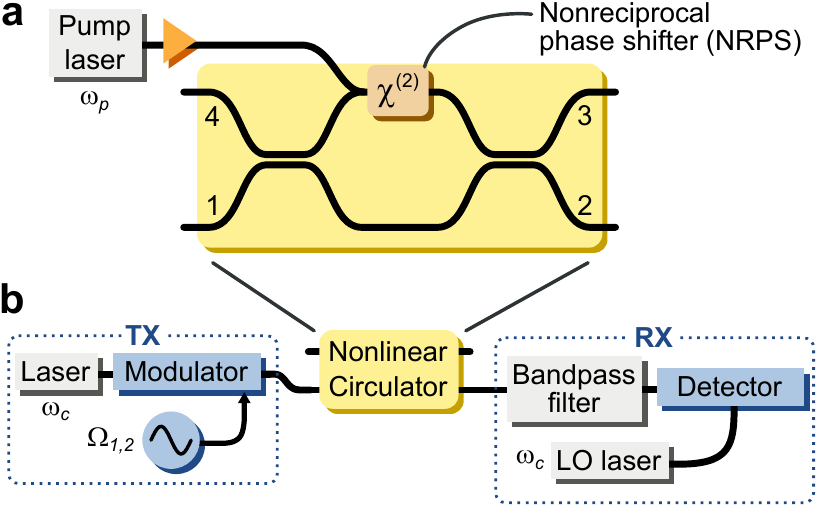}
  \caption{(a) Schematic of a parametric optical circulator implemented with a waveguide nonreciprocal phase shifter (NRPS) in one arm of a dual-input dual-output Mach-Zehnder interferometer (DIDO-MZI). The NRPS is achieved in the signal-idler Rabi oscillation of a $\chi^{(2)}$ nonlinear waveguide. (b) Schematic of a microwave photonic link transiting the parametric nonlinear circulator. A bandpass filter allows only the signal waves to reach the photo detector, while eliminating the pump and any residual idler waves.}
  \label{fig:fig1}
\end{figure}

In this letter we first review the operating principals of a parametric nonlinear optical circulator. We then numerically characterize the broadband performance of the circulator by modeling its response to a two-tone signal, allowing us to characterize intermodulation distortion, dynamic range, and the allowable signal power levels of the circulator relative to the pump wave. Our results reveal that that despite not being constrained by dynamic reciprocity, parametric nonlinear circulators suffer significant signal distortion and a reduced dynamic range even in the undepleted pump regime. This could potentially inhibit their use in next generation optical communications and signal processing systems requiring very high dynamic range.

Parametric nonlinear circulators are implemented in resonant \cite{hua_demonstration_2016} or waveguiding geometries \cite{yu_complete_2009}, where photonic modes are coupled via symmetry and momentum matching of the modulating radio frequency signal or the optical pump. Here we focus only on a waveguide implementation that leverages a nonreciprocal phase shift (NRPS). When inserted into one arm of a dual-input dual-output Mach-Zender interferometer (DIDO-MZI), as shown in Fig. \ref{fig:fig1}a, the NRPS is converted into a circulator response in terms of scattered power, with transmission occurring from port 1 $\rightarrow$ port 2 $\rightarrow$ port 4 $\rightarrow$ port 3 $\rightarrow$ port 1.

The NRPS in the parametric circulator is achieved through a $\chi^{(2)}$ Rabi oscillation between signal and idler waves within a waveguide \cite{wang_non-reciprocal_2017}. In the undepleted pump limit, where the pump amplitude is much higher than the signal and idler amplitude, $A_p \gg A_{s(i)}$ and with perfect phase matching, the amplitude of the signal and idler waves have a sinusoidal spatial dependence \cite{boyd_nonlinear_2008},
\begin{align}
A_s\left(z\right) &= A_s\left(0\right)\text{cos}\left(\kappa z\right) \label{eq:analytic1} \\
A_i\left(z\right) &= A_s\left(0\right) j\sqrt{\frac{n_s\omega_i}{n_i\omega_s}}e^{j\phi_p}\text{sin}\left(\kappa z\right).
\label{eq:analytic2}
\end{align}
In the more general case of the signal wave amplitude approaching the pump amplitude, Jacobi elliptic functions describe the evolution of the wave amplitudes \cite{armstrong_interactions_1962}. The nonlinear coupling coefficient, 
\begin{equation}
\kappa = 2\frac{\omega_s \omega_i d_{\text{eff}}}{\sqrt{k_sk_i}c^2}\vert A_p\vert
\end{equation}
defines the characteristic interaction length of the parametric process. From Eqn. \ref{eq:analytic1} it is clear that undergoing one Rabi cycle, $\kappa z = \pi$ results in a $\pi$ phase shift for the signal wave. Moreoever, this phase is nonreciprocal or directional via the momentum matching condition of the parametric process, $k_s+k_p=k_i$ through the engineered dispersion of an on-chip waveguide, and can also be viewed as an indirect mode transition induced by the pump \cite{yu_complete_2009}.

Although useful for highlighting the operating principals of the nonlinear waveguide, the transfer function for the signal described by Eqn. \ref{eq:analytic1} does not apply to the case of a finite-bandwidth input signal. In general, the broadband response is not simply the superposition of the response at individual frequencies due to cascaded mixing among different tones that make up the signal and idler waves \cite{barbour_high_2016}. Parasitic multi-tone mixing leeches energy from the desired signal $\rightarrow$ idler $\rightarrow$ signal process, and generates spurious tones that interfere with the signals being routed by the circulator. 

\begin{figure}
  \centering
  \includegraphics{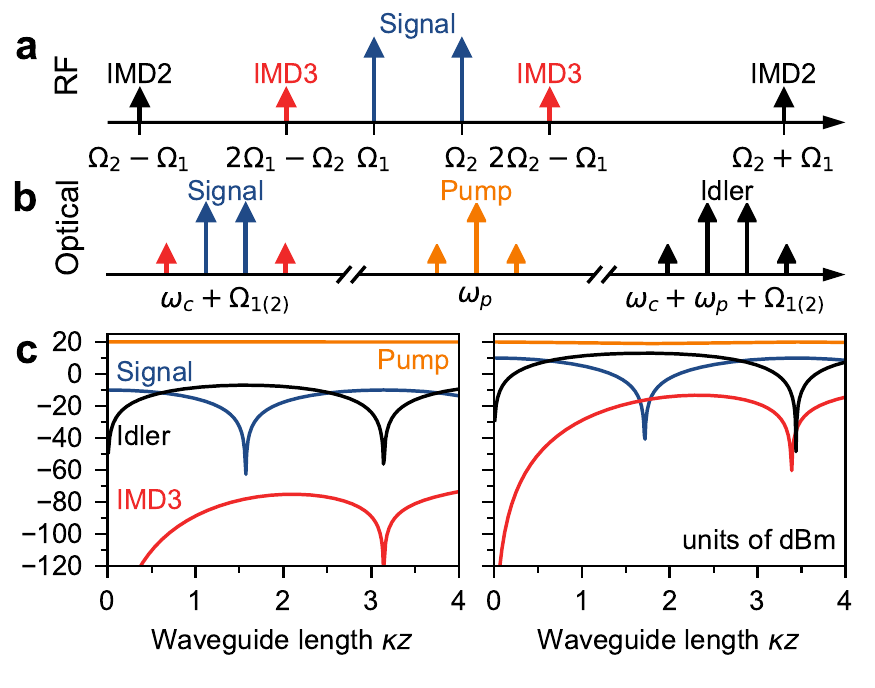}
  \caption{(a) RF spectrum showing two signal tones at $\Omega_1$ and $\Omega_2$ along with the second-order and third-order intermodulation tones (IMD2 and IMD3). (b) Corresponding optical tones after upconversion. (c) Evolution of the pump, signal, idler, and IMD3 signal spurs along the length $z$ of the $\chi^{(2)}$ waveguide for input signal tone powers of -10 dBm and +10 dBm. The pump power is 20 dBm. The horizontal axis is normalized to the pump wave amplitude and the nonlinear coupling coefficient. Revival of the signal wave occurs for $z \kappa = \pi$.}
  \label{fig:fig2}
\end{figure}

In microwave photonics the linearity of an optical link, in terms of a fundamental input and output RF signal, is characterized with a \textit{two-tone} test \cite{urick_fundamentals_2015}. This underpins the system's response to a broadband excitation, where the two tones represent any arbitrary pair of frequencies within the continuous input spectrum. A test setup for exciting two optical tones is the single-sideband (SSB) transmitter architecture and direct detection receiver (Fig. \ref{fig:fig1}b). The two fundamental microwave tones are $\Omega_1$ and $\Omega_2$ which are upconverted to $\omega_c+\Omega_{1,(2)}$, where $\omega_c$ is the optical carrier (Fig. \ref{fig:fig2}a,b). In this architecture the optical carrier at $\omega_c$ is filtered at the output of the modulator. Intermodulation distortion (IMD) products arise in the output microwave spectrum at combinations of the two fundamental input tones (Fig. \ref{fig:fig2}a): second order intermodulation distortion (IMD2) spurs occur at frequencies $\Omega_2\pm\Omega_1$ and third order intermodulation distortion (IMD3) spurs occur at frequencies $2\Omega_{1(2)}-\Omega_{2(1)}$.

In the nonlinear parametric phase shifter, the third order spur (IMD3) dominates as the result of a three-stage cascaded nonlinear process within the waveguide. First, the two input signal tones are upconverted to tones on the idler wave at $\omega_c+\omega_p+\Omega_{1(2)}$. In the second stage, each idler tone at $\omega_c+\omega_p+\Omega_{1(2)}$ mixes with its opposite signal tone at $\omega_c+\Omega_{2(1)}$ to generate spur tones around the pump wave at $\omega_p+\Omega_{1(2)}-\Omega_{2(1)}$. Finally, the sprs around the pump wave mix with the signal (idler) tone in a SFG (DFG) process to generate the spurs around the idler (signal) tones. These spurs around the signal tone ultimately show up as IMD3 tones in the output microwave spectrum after demodulating from the optical carrier.

The proximity of these IMD3 spurs to the signal tones inherently limits the signal fidelity because they can not be optically filtered. Moreover, because the incident signals at any of the MZI ports pass through the $\chi^{(2)}$ waveguide (Fig. \ref{fig:fig1}a), the resulting IMD3 spurs will have the potential to interfere not only within their own pathway, but also with other pathways in the circulator. For example, the IMD3 spurs generated from the signal incident at port 1 will show up in the output of both port 2 and port 3. Similarly to the limitation of dynamic reciprocity \cite{shi_limitations_2015}, this interference presents a fundamental challenge for full-duplex communications where a high-power transmit signal must be routed simultaneously with a much lower power received signal. The IMD3 spurs resulting from the high power transmit signal would completely overwhelm the much lower power received signals.

The output power of the IMD3 spur ultimately determines the operating signal power levels of the circulator. As a concrete example, a pump power of 20 dBm with an input signal power of -10 dBm (in both tones) generates an output IMD3 spur that is less than -120 dBm at the waveguide length for complete signal revival, $z \kappa = \pi$ (Fig. \ref{fig:fig2}c, left side). However, when the input power of the two signal tones is increased to 10 dBm, the IMD3 spur power increases to approximately -40 dBm for the same waveguide length (Fig. \ref{fig:fig2}c, right side). This is due to a nonlinear shift in the spatial null of the IMD3 spur which arises from the cascaded nonlinear process described above. These results were computed by numerically solving the nonlinear coupled amplitude equations \cite{gallo_efficient_1997, chen_analysis_2004}, accounting for all frequency components shown in Fig. \ref{fig:fig2}b. The template coupled amplitude equation for a wave with index $l$ is 
\begin{equation}
\begin{split}
\frac{dA_l}{dz} =\sum_{m,n}j\frac{1}{2}\frac{d_{\text{eff}}\omega_l^2}{k_l c^2} A_m A_n e^{j\left(k_l-k_n-k_m\right)z} \delta\left(\omega_l-\omega_n-\omega_m\right) \\
+\sum_{m,n}j\frac{d_{\text{eff}} \omega_l^2}{k_l c^2} A_m A_n^* e^{j\left(k_l+k_n-k_m\right)z} \delta\left(\omega_l+\omega_n-\omega_m\right).
\end{split}
\label{eq:template}
\end{equation}
In Eqn. \ref{eq:template}, the first term captures all possible sum frequency generation (SFG) processes that couple with the wave at $\omega_l$, and the second term captures all possible difference frequency generation (DFG) processes that couple with the wave at $\omega_l$.

\begin{figure}[t]
  \centering
  \includegraphics{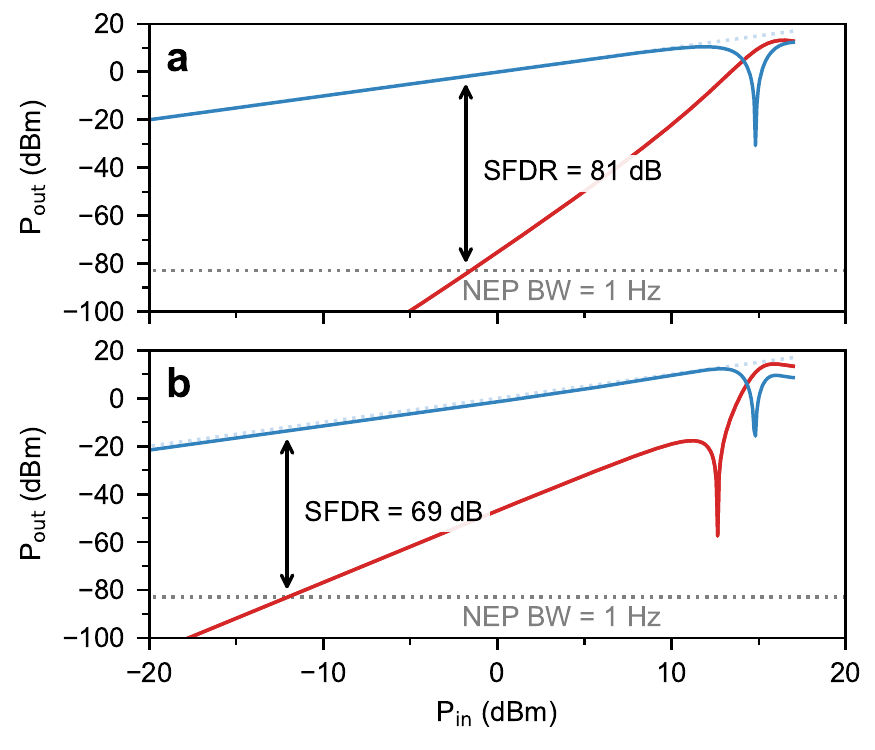}
  \caption{Output power of fundamental signal tones (blue) and third-order intermodulation distortion products (IMD3) (red) as a function of the input signal power for a normalized waveguide length of (a) $z \kappa = \pi$ and (b) $z \kappa = 1.184\pi$. The pump power $P_p$ = 20 dBm and the optical noise equivalent power is $P_{\text{NEP}}$ = -83 dBm.}
  \label{fig:fig3}
\end{figure}

The linearity of the signal transfer function holds for power levels well below the pump power where a waveguide length satisfying $z \kappa = \pi$, completely recovers both signal tones with no insertion loss (Fig. \ref{fig:fig3}a). Here both signal tones have identical input powers and have essentially the same same spatial dependence along the nonlinear waveguide. For signal powers that are within 8 dB of the pump, compression of the output signal tones occurs. At such high signal power, the pump wave can no longer provide enough energy to sustain the Rabi oscillation between the idler and signal waves, which is critical for achieving a linear response in the NRPS. At first glance, compression may be seen as the upper limit on allowable signal power but in fact, the IMD3 spur limits the system response before the onset of pump depletion. A noise floor is used to quantify the relative strength of the intermodulation spurs. In the optical domain the photo diode noise equivalent power (NEP) is defined as the optical intensity that results in a signal-to-noise ratio (SNR) of unity in the output, $P_\text{NEP} = S \sqrt{B}/\mathcal{R}$ = -83 dBm. Here, we assume a photodiode noise output spectral density $S$ = 0.6 A/$\sqrt{\text{Hz}}$, a photodiode responsivity $\mathcal{R}$ = 0.8 A/W, and a detector bandwidth $B$ = 1 Hz. 

The input intercept point (IIP) and output intercept point (OIP) define the input and output signal power, respectively, where the system produces an equal signal and spur power. These points are used to extrapolate back to the noise floor and calculate a spur-free dynamic range (SFDR) as
\begin{equation}
\text{SFDR}_n = \frac{n-1}{n}\left( \text{IIP}_n - P_{\text{noise}} \right),
\label{eq:sfdr}
\end{equation}
where $n = \left\{ 2,3,\ldots\right\}$ is the order of the dominant intermodulation spur. At the length for complete revival of the signal wave $z\kappa=\pi$, the SFDR reaches its maximal value of 81 dB due to the spatial null in the IMD3 spur for low signal powers (Fig. \ref{fig:fig3}a). For a perturbed length (or pump power) where $z\kappa\approx3.7120 = 1.184\pi$, the SFDR drops to 69 dB due operation away from the spatial null of the IMD3 spur (Fig. \ref{fig:fig3}b). This indicates that obtaining the largest possible SFDR requires careful balancing of the system fabrication constraints as well as the pump power that is ultimately coupled into the on-chip nonlinear waveguide. At the waveguide length and pump power with maximal SFDR $\left(z\kappa=\pi\right)$, the IMD3 slope is 6 in log-log scaling, rather than 3, even though the spur is third-order. The much steeper dependence on input signal power at this length is what opens up a wider spur-free range of operation.

\begin{figure}[t]
  \centering
  \includegraphics{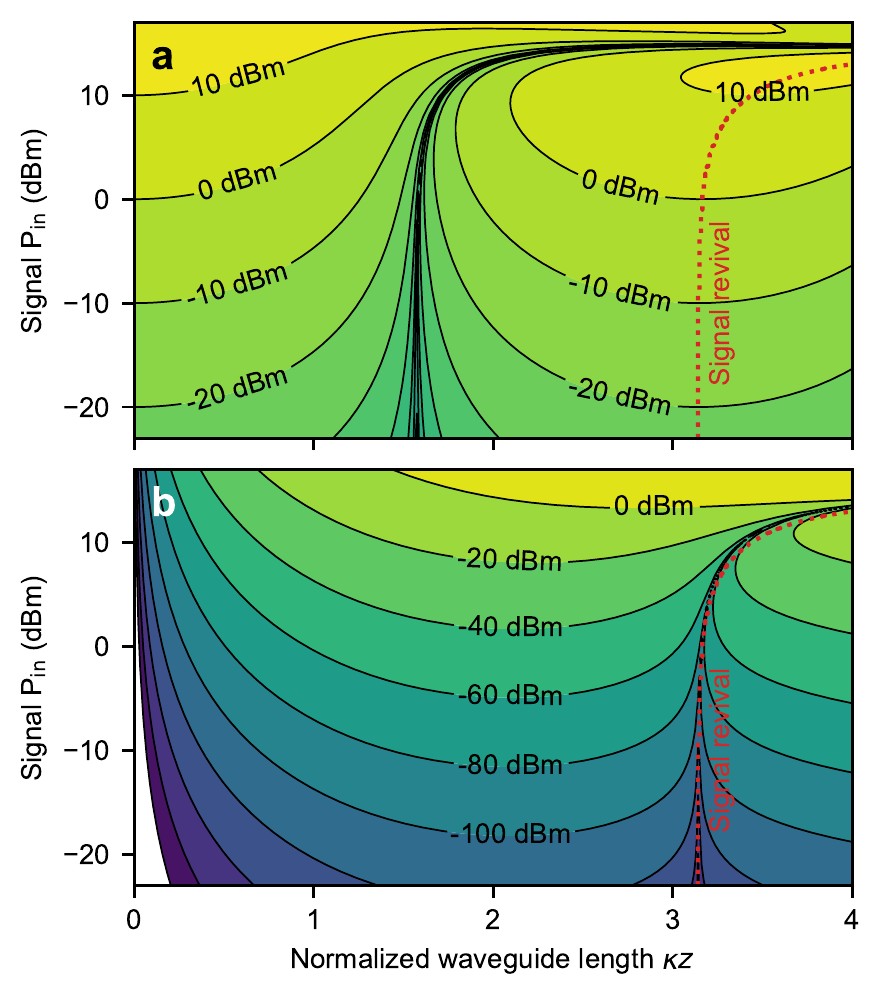}
  \caption{Design chart for a nonlinear optical phase shifter showing output power levels of (a) signal waves at $\omega_c+\Omega_{1(2)}$ and (b) third-order intermodulation distortion products (IMD3) at $\omega_c+2\Omega_{1(2)}-\Omega_{2(1)}$. The vertical axis is the input signal power and the horizontal axis is the normalized length of the waveguide. The pump power $P_p$ = 20 dBm and the red dashed contour indicates complete recovery of the signal waves and begins to curve at high input powers due to pump depletion.}
  \label{fig:fig4}
\end{figure}

The maximum SFDR of 81 dB is far below the SFDR of modern microwave optical links which can be on the order of 110 dB or greater \cite{marpaung_integrated_2013}. This points to this particular parametric circulator architecture being a bottleneck for signal fidelity. The computed design chart showing the signal power level as a function of both waveguide length and input power shows the fundamental challenge of simultaneously maintaining low insertion loss, avoiding signal compression, and having a high dynamic range (Fig. \ref{fig:fig4}a).

Only a narrow region of the parameter space can be used for optimal device operation, where the dotted red contour indicates complete signal revival. Its curvature at large signal powers indicates that for a fixed waveguide length and pump power, negligible insertion loss can not be achieved for all possible input signal powers. Complete revival can occur for a single waveguide length, but only up to powers of approximately 5 dBm, which is 15 dB below the pump. The null in the IMD3 spur closely tracks the signal revival contour but diverges slightly once the input signal power exceeds approximately 5 dBm (Fig. \ref{fig:fig4}b). The corresponding null is where the maximum SFDR is obtained. On the other hand, operating at a waveguide length and pump power where $z\kappa\ne\pi$, doesn't result in a significant increase in the link's insertion loss. However, a much more significant penalty is observed in the SFDR as is also confirmed by Fig. \ref{fig:fig3}a.

Our numerical modeling indicates that despite not being constrained by dynamic reciprocity, nonlinear parametric optical circulators have a reduced dynamic range that may be too low for next generation all-optical signal processing systems. In order to improve the dynamic range from the 81 dB predicted by our simulations, either the noise floor must be reduced or the pump power must be increased. For a lower noise floor, the increase in dynamic range will be associated with a reduced maximum signal power, corresponding to the intersection of the dashed horizontal line and the red spur line of Fig. \ref{fig:fig3}a,b. Importantly, the intermodulation distortion resulting from the NRPS will carry over to adjacent ports (e.g port 3, when transmitting from port 1 to port 2) because the spurs from the $\chi^{(2)}$ process can not be canceled at the 50/50 coupler. In practice, this will result in significant in-band interference between simultaneous excitations of the circulator ports, such as a full-duplex application. Although we have only considered a nonlinear waveguide geometry in this work, we believe that similar conclusions will apply to resonant nonlinear processes. The linewidth of an optical cavity can not filter IMD3 spurs generated from arbitrarily small signal tones.

\section*{Methods}
In this work the effective nonlinear coupling was $d_{\text{eff}} = 2.5 \times 10^{-12}$ V/m, the effective mode area was $S = 1 \times 1$ $\mu$m$^2$, the waveguide length was 50.638 mm for $\kappa z = 4$, and the pump power was 20 dBm. The RF signal tones were $\Omega_1$ = 9 MHz and $\Omega_2$ = 11 MHz. The optical carrier was $\omega_c$ = 200.1 THz and the pump frequency was $\omega_p$ = 201.2 THz. All results in this work were computed by expanding the template Eqn. \ref{eq:template} into a set accounting for all 11 frequencies shown in Fig. \ref{fig:fig3}b with the sum- and difference-frequency coupling accounted for by the Dirac delta functions. The resulting set of coupled equations was solved with Newton's method and a Crank-Nicolson finite differencing scheme \cite{crank_practical_1996}.

\section*{Acknowledgments}
This work was supported by the Packard Fellowships for Science and Engineering, the National Science Foundation (NSF) (EFMA-1641069), and the US Office of Naval Research (ONR) (N00014-16-1-2687).

\bibliography{refs}

\end{document}